\documentstyle[12pt]{article}

\catcode`@=11
\def\un#1{\relax\ifmmode\@@underline#1\else
        $\@@underline{\hbox{#1}}$\relax\fi}
\catcode`@=12

\def\a{\alpha}
\def\b{\beta}
\def\c{\chi}
\def\d{\delta}

\def\f{\phi}
\def\g{\gamma}
\def\h{\eta}

\def\j{\psi}

\def\l{\lambda}
\def\m{\mu}
\def\n{\nu}

\def\p{\pi}

\def\s{\sigma}
\def\t{\tau}

\def\x{\xi}
\def\z{\zeta}

\def\L{\Lambda}

\def\ve{\varepsilon}

\def\co{{\cal O}}

\def\bo{{\raise-.5ex\hbox{\large$\Box$}}}               
\def\pa{\partial}                                       
\def\TH{{\raise.2ex\hbox{$\displaystyle \bigodot$}\mskip-4.7mu \llap H \;}}
\def\face{{\raise.2ex\hbox{$\displaystyle \bigodot$}\mskip-2.2mu \llap {$\ddot
        \smile$}}}                                      

\def\sp#1{{}^{#1}}                              
\def\Hat#1{\widehat{#1}}                        
\def\leftrightarrowfill{$\mathsurround=0pt \mathord\leftarrow \mkern-6mu
        \cleaders\hbox{$\mkern-2mu \mathord- \mkern-2mu$}\hfill
        \mkern-6mu \mathord\rightarrow$}
\def\dvec#1{\vbox{\ialign{##\crcr
        \leftrightarrowfill\crcr\noalign{\kern-1pt\nointerlineskip}
        $\hfil\displaystyle{#1}\hfil$\crcr}}}           

\def\frac#1#2{{\textstyle{#1\over\vphantom2\smash{\raise.20ex
        \hbox{$\scriptstyle{#2}$}}}}}                   
\def\ha{\frac12}                                        
\def\sfrac#1#2{{\vphantom1\smash{\lower.5ex\hbox{\small$#1$}}\over
        \vphantom1\smash{\raise.4ex\hbox{\small$#2$}}}} 
\def\bfrac#1#2{{\vphantom1\smash{\lower.5ex\hbox{$#1$}}\over
        \vphantom1\smash{\raise.3ex\hbox{$#2$}}}}       
\def\afrac#1#2{{\vphantom1\smash{\lower.5ex\hbox{$#1$}}\over#2}}    

\def\[{\lfloor{\hskip 0.35pt}\!\!\!\lceil}
\def\]{\rfloor{\hskip 0.35pt}\!\!\!\rceil}

\def\ud#1#2{^{#1}{}_{#2}}

\def\fracm#1#2{\hbox{\large{${\frac{{#1}}{{#2}}}$}}}
\def\half{{\fracm12}}
\def\ha{\half}
\def\tr{{\rm tr}}

\def\un{\underline}
\def\fracmm#1#2{{{#1}\over{#2}}}

\def\low#1{{\raise -3pt\hbox{${\hskip 0.75pt}\!_{#1}$}}}

\def\Hat#1{\widehat{#1}}

\newskip\humongous \humongous=0pt plus 1000pt minus 1000pt
\def\caja{\mathsurround=0pt}
\def\eqalign#1{\,\vcenter{\openup2\jot \caja
        \ialign{\strut \hfil$\displaystyle{##}$&$
        \displaystyle{{}##}$\hfil\crcr#1\crcr}}\,}
\newif\ifdtup

\def\ref#1{$\sp{#1)}$}

\def\pl#1#2#3{Phys.~Lett.~{\bf {#1}B} (19{#2}) #3}
\def\np#1#2#3{Nucl.~Phys.~{\bf B{#1}} (19{#2}) #3}

\parskip=\medskipamount                 
\thicklines                         

\begin{document}


\thispagestyle{empty}               

\def\border{                                            
        \setlength{\unitlength}{1mm}
        \newcount\xco
        \newcount\yco
        \xco=-24
        \yco=12
        \begin{picture}(140,0)
        \put(-20,11){\tiny Institut f\"ur Theoretische Physik Universit\"at
Hannover~~ Institut f\"ur Theoretische Physik Universit\"at Hannover~~
Institut f\"ur Theoretische Physik Hannover}
        \put(-20,-241.5){\tiny Institut f\"ur Theoretische Physik Universit\"at
Hannover~~ Institut f\"ur Theoretische Physik Universit\"at Hannover~~
Institut f\"ur Theoretische Physik Hannover}
        \end{picture}
        \par\vskip-8mm}

\def\headpic{                                           
        \indent
        \setlength{\unitlength}{.8mm}
        \thinlines
        \par
        \begin{picture}(29,16)
        \put(75,16){\line(1,0){4}}
        \put(80,16){\line(1,0){4}}
        \put(85,16){\line(1,0){4}}
        \put(92,16){\line(1,0){4}}

        \put(85,0){\line(1,0){4}}
        \put(89,8){\line(1,0){3}}
        \put(92,0){\line(1,0){4}}
        \put(85,0){\line(0,1){16}}
        \put(96,0){\line(0,1){16}}
        \put(79,0){\line(0,1){16}}
        \put(80,0){\line(0,1){16}}
        \put(89,0){\line(0,1){16}}
        \put(92,0){\line(0,1){16}}
        \put(79,16){\oval(8,32)[bl]}
        \put(80,16){\oval(8,32)[br]}

        \end{picture}
        \par\vskip-6.5mm
        \thicklines}

\border\headpic {\hbox to\hsize{
\vbox{\noindent ITP--UH--13/94 \\ hep-th/9409020
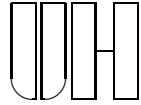
\hfill September 1994}}}

\noindent
\vskip1.3cm
\begin{center}

{\Large\bf How Many $N=4$ Strings Exist ?}
\footnote{Supported in part by the `Deutsche Forschungsgemeinschaft' and the
NATO grant CRG 930789}\\
\vglue.3in

Sergei V. Ketov \footnote{On leave of absence from:
High Current Electronics Institute of the Russian Academy of Sciences,
Siberian Branch, Akademichesky~4, Tomsk 634055, Russia}

{\it Institut f\"ur Theoretische Physik, Universit\"at Hannover}\\
{\it Appelstra\ss{}e 2, 30167 Hannover, Germany}\\
{\sl ketov@kastor.itp.uni-hannover.de}
\end{center}
\vglue.2in
\begin{center}
{\Large\bf Abstract}
\end{center}
Possible ways of constructing extended fermionic strings with $N=4$ world-sheet
 supersymmetry are reviewed. String theory constraints form, in general, a
non-linear quasi(super)conformal algebra, and can have conformal dimensions
$\geq 1$. When $N=4$, the most general $N=4$ quasi-superconformal algebra to
consider for string theory building is $\hat{D}(1,2;\a)$, whose linearisation
is the so-called `large' $N=4$ superconformal algebra. The $\hat{D}(1,2;\a)$
algebra has $\Hat{su(2)}_{k^+}\oplus \Hat{su(2)}_{k^-}\oplus\Hat{u(1)}$
Ka\v{c}-Moody component, and $\a=k^-/k^+$. We check the Jacobi identities and
construct a BRST charge for the $\hat{D}(1,2;\a)$ algebra. The quantum BRST
operator can be made nilpotent only when $k^+=k^-=-2$. The $\hat{D}(1,2;1)$
algebra is actually isomorphic to the $SO(4)$-based Bershadsky-Knizhnik
non-linear quasi-superconformal algebra. We argue about the existence of a
string theory associated with the latter, and propose the (non-covariant)
hamiltonian action for this new $N=4$ string theory. Our results imply the
existence of two different $N=4$ fermionic string theories: the old
one based on the `small' linear $N=4$ superconformal algebra and having the
total ghost central charge $c_{\rm gh}=+12$, and the new one with non-linearly
realised $N=4$ supersymmetry, based on the $SO(4)$ quasi-superconformal algebra
and having $c_{\rm gh}=+6$. Both critical string theories have negative
`critical dimensions' and do not admit unitary matter representations.

\newpage
\hfuzz=10pt

\section{Introduction}

Any known critical $N$-extended fermionic string theory with $N\leq 4$
world-sheet supersymmetries is based on a two-dimensional (2d) linear
$N$-extended {\it superconformal algebra} (SCA) which is gauged \cite{aba}.
The string world-sheet fields usually form a linear $N$-extended superconformal
 multiplet coupled to the $N$-extended conformal supergravity fields which are
the gauge fields of the $N$-extended SCA. This pattern worked well for the
$N=1$ and $N=2$ strings, and it was always expected to be also true for $N=4$
strings. In the past, only one $N=4$ string theory had been actually
constructed \cite{aba2} by gauging the `small' linear $N=4$ SCA with $SU(2)$
internal symmetry. The 2d covariant action of this $N=4$ string can be found
in ref.~\cite{pn}. An apparent drawback of this $N=4$ string theory is the
 `negative' value of its (quaternionic) critical dimension, $D_{\rm c}=-2$,
which always prevented this theory from having physical applications. This
explains, in particular, why nobody succeeded in constructing $N=4$ string
scattering amplitudes.

Still, it is of interest to know how many different $N=4$ string theories can
be constructed at all. Any $N=4$ string constraints are going to be very
strong in two dimensions, so that their explicit realisation should always
imply very non-trivial interplay between geometry, conformal invariance and
extedned supersymmetry. The $N=4$ strings are also going to be relevant in the
search for the `universal string theory' \cite{bvu}. In addition, strings with
$N=4$ supersymmetry are expected to have deep connections with integrable
models \cite{ket,book}, so that we believe they are worthy to be studied.

It has been known for some time that there are actually {\it two} different
linear $N=4$ SCAs which are (affine verions of) finitely-generated Lie
superalgebras: the so-called `{\it small}' linear $N=4$ SCA with $SU(2)$
internal symmetry \cite{aba,aba2}, and the so-called `{\it large}' linear $N=4$
 SCA with $SU(2)\times SU(2)\times U(1)$ internal symmetry \cite{sch1,stvp}.
Unlike the `small' $N=4$ SCA mentioned above, the `large' $N=4$ SCA has
{\it subcanonical} charges, or currents of conformal dimension $1/2$. This
observation already implies that no supergravity or string theory based on the
 `large' $N=4$ SCA exists, because there are no 2d gauge fields which would
correspond to the fermionic charges of dimension $1/2$.~\footnote{In conformal
field theory, `currents' of dimension $1/2$ are just free fermions
\cite{book}.}

When a number of world-sheet supersymmetries exceeds two, there are, in fact,
more opportunities to build up new string theories, by using 2d
{\it non-linear} quasi-superconformal algebras which are known to exist for an
arbitrary $N>2$. By an $N$-extended {\it quasi-superconformal algebra} (QSCA)
we mean a graded associative algebra whose contents is restricted to canonical
charges of dimension $2$, $3/2$ and $1$, which (i) contains the Virasoro
subalgebra, and (ii) $N$ real supercurrents of conformal dimension $3/2$,
whose {\it operator product expansion} (OPE) has a stress tensor of dimension
$2$, (iii) satisfies the Jacobi identity, and (iv) has the usual
spin-statistics relation.~\footnote{We exclude from our analysis all kinds of
{\it twisted} (Q)SCAs with unusual relations between spin and statistics (they
are, however, relevant for topological field theory and topological strings
\cite{bv}). We also ignore all kinds of reducible combinations of SCAs and
QSCAs.} By definition, a QSCA is an `almost' usual SCA, except it may not be a
Lie superalgebra but its OPEs have to be closed on quadratic composites of the
fundamental set of canonical generators. The QSCAs can, therefore, be
considered on equal footing with the $W$ algebras \cite{bs} {\it without},
however, having currents of spin higher than two. Though QSCAs do not belong,
in general, to ordinary (finitely-generated) affine Lie superalgebras, but, so
to say, to infinitely-generated Lie superalgebras, they are still closely
related with finite Lie superalgebras~\cite{fl}.

The full classification of QSCAs has been done by Fradkin and Linetsky
\cite{fl}.~\footnote{See also refs.~\cite{rs,htt} for earlier works on
classification of linear SCAs.} The classification \cite{fl} is based on the
classical results of Ka\v{c} \cite{kac} on classification of finite simple Lie
superalgebras, while their  relation to QSCAs is very similar to the
one existing between ordinary Lie algebras $sl(n)$ and $W_n$ algebras via a
Drinfeld-Sokolov-type reduction \cite{bs}. There are three classical families
of QSCAs,  namely
$$osp(N|2;{\bf R})~,\quad su(1,1|M)~,\quad osp(4^*|2M)~,$$
one {\it continuous} family $\hat{D}(2,1;\a)$ parametrised by a real parameter
$\a$,~\footnote{The algebras $\hat{D}(2,1;\a)$ and $\hat{D}(2,1;\a^{-1})$ are
isomorphic.} and the two exceptional QSCAs with $N=7$ and $N=8$ supersymmetry
 \cite{fl,bo}.

When $N=4$, the only different QSCAs are just $su(1,1|2)$ and
$\hat{D}(2,1;\a)$,
since $osp(4^*|2)$ is isomorphic to $osp(4|2;{\bf R})$, whereas the latter
 can actually be considered as a particular member of the family
$\hat{D}(2,1;\a)$,
as we are going to demonstrate in this paper. The $su(1,1|2)$ QSCA is, in fact,
a linear $N=4$ SCA which is isomorphic to the `small' $SU(2)$-based $N=4$ SCA.

The $osp(N|2;{\bf R})$ and $su(1,1|N)$ series of QSCAs with the $\Hat{SO(N)}$
and $\Hat{U(N)}$ {\it Ka\v{c}-Moody} (KM) symmetries, respectively, were
discovered by Knizhnik \cite{kn} and Bershadsky \cite{be}, whereas the
non-linear $\hat{D}(2,1;\a)$ QSCA was originally extracted by Goddard and
Schwimmer \cite{gs} from the `large' linear $N=4$ SCA of ref.~\cite{sch1,stvp}
by factoring out free fermions and boson. In our paper, we give a straitforward
construction of this QSCA, and emphasize on some peculiarities of the $N=4$
case in the orthogonal (Bershadsky-Knizhnik) series of the QSCAs.

Since the $N=4$ supersymmetric $\hat{D}(2,1;\a)$ QSCA includes only canonical
charges, despite of its apparent non-linearity it seems possible to construct
the associated 2d, $N=4$ conformal supergravity and, hence, a new $N=4$
string theory by coupling this supergravity to an appropriate 2d matter,
completely along the lines of constructing the $W$ gravities and $W$ strings.

The paper is organised as follows. In sect.~2 we review the Bershadsky-Knizhnik
 orthogonal series of non-linear $N$-extended QSCAs in two dimensions, paying
special attention to a peculiar nature of the $N=4$ case. In sect.~3, the
$N=4$ extended nonlinear  $\hat{D}(1,2;\a)$ QSCA is introduced, and its
relation to
the `large' linear $N=4$ SCA is explained. In sect.~4 we discuss a construction
 of {\it classical} superconformal field theories and supergravities, in which
quasi(super)conformal algebras appear as symmetry algebras, since we are
interested in {\it local} realisations of the corresponding symmetries, needed
for supergravity and string theory building. The hamiltonian (2d non-covariant)
form of invariant action is also presented in sect.~4. The BRST quantisation
and the construction of quantum BRST charges for both the $SO(N)$-based
Bershadsky-Knizhnik QSCA and the $\hat{D}(1,2;\a)$ QSCA are given in sect.~5.
Sect.~6 comprises our conclusion. In Appendix, the relevant facts, needed for a
construction of the classical BRST charge in a gauge theory with first-class
constraints satisfying a non-linear algebra, are summarized.
\vglue.2in

\section{Bershadsky-Knizhnik orthogonal QSCA series}

The current contents of the 2d $N$-extended Bershadsky-Knizhnik QSCA
\cite{kn,be} is given by the holomorphic fields $T(z)$, $G^i(z)$ and
$J^a(z)$, all having the standard mode expansions
$$\eqalign{
T(z)= & \sum_n L_n z^{-n-2}~,\cr
G^i(z)= & \sum_r G^i_r z^{-r-3/2}~,\cr
J^a(z)= & \sum_n J^a_n z^{-n-1}~,\cr}\eqno(2.1)$$
and (conformal) dimensions $2$, $3/2$ and $1$, respectively. The real
supersymmetry generators $G^i$, $i=1,\ldots,N$, are defined in the fundamental
(vector) representation of the internal symmetry group $SO(N)$ generated by
the zero modes of the currents $J^a$, $a=1,\ldots,{\frac 12}N(N-1)$, in
the adjoint representation.

Most of the OPEs defining the $SO(N)$-based Bershadsky-Knizhnik QSCA take the
standard linear form, {\it viz.}
$$\eqalign{
T(z)T(w)~\sim~ & \fracmm{c/2}{(z-w)^4} + \fracmm{2T(w)}{(z-w)^2}
+ \fracmm{\pa T(w)}{z-w}~,\cr
T(z)G^i(w)~\sim~ & \fracmm{{\frac 32}G^i(w)}{(z-w)^2} +
 \fracmm{\pa G^i(w)}{z-w}~,\cr
T(z)J^a(w)~\sim~ & \fracmm{J^a(w)}{(z-w)^2} + \fracmm{\pa J^a(w)}{z-w}~,\cr
J^a(z)G^i(w)~\sim~ & \fracmm{(t^a)^{ij}G^j(w)}{z-w}~,\cr
J^a(z)J^b(w)~\sim~ & \fracmm{f^{abc}J^c(w)}{z-w}
+ \fracmm{-k\d^{ab}}{(z-w)^2}~,\cr}\eqno(2.2)$$
where $k$ is an arbitrary `level' of the KM subalgebra, $f^{abc}$ are $SO(N)$
structure constants, and $(t^a)^{ij}$ are generators of $SO(N)$ in the
fundamental (vector) representation,
$$\eqalign{
\[t^a,t^b\]=f^{abc}t^c,\quad & \quad  f^{abc}f^{abd}=2(N-2)\d^{cd}~,\cr
\tr(t^at^b)=-2\d^{ab}~,\quad & \quad
 (t^a)^{ij}(t^a)^{kl}=\d^{ik}\d^{jl}-\d^{il}\d^{jk}~.\cr}\eqno(2.3)$$

On symmetry and dimensional reasons, the only non-trivial OPE defining the
non-linear supersymmetry component of QSCA can be of the form
$$\eqalign{
G^i(z)G^j(w) ~\sim~ & a_1\fracmm{\d^{ij}}{(z-w)^3} + a_2\fracmm{(t^a)^{ij}
J^a(w)}{(z-w)^2} +\fracmm{1}{z-w}\left[ 2\d^{ij}T(w)
+ \ha a_2(t^a)^{ij}\pa J^a(w)\right]\cr
& +\fracmm{1}{z-w}\left[ a_3\left(t^{(a}t^{b)}\right)^{ij}
+a_4\d^{ab}\d^{ij}\right]:J^aJ^b:(w)~,\cr}\eqno(2.4)$$
where $a_1,\,a_2,\,a_3$ and $a_4$ are parameters to be determined by
solving the Jacobi identity, and the normal ordering is defined by
$$:J^aJ^b:(w)=\lim_{z\to w}\left[J^{(a}(z)J^{b)}(w) +
 \fracmm{k\d^{ab}}{(z-w)^2}\right]~.\eqno(2.5)$$
Indices in brackets mean symmetrization with unit weight,
e.g. $t^{(a}t^{b)}\equiv {\frac 12}(t^at^b+t^bt^a)$. Eq.~(2.4) can be
considered as the general {\it ansatz} for supersymmetry algebra.

Demanding consistency of the whole algebra determines the parameters
\cite{kn,be}:
$$\eqalign{
a_1=\fracmm{k(N+2k-4)}{N+k-3}~,\quad & \quad a_2= \fracmm{N+2k-4}{N+k-3}~,
\cr
a_3=  a_4 = \fracmm{1}{N+k-3}~,\quad & \quad }\eqno(2.6)$$
while the Virasoro central charge of this QSCA is also quantized \cite{kn,be}:
$$ c = \fracmm{k(N^2+6k-10)}{2(N+k-3)}~.\eqno(2.7)$$
The KM parameter $k$ remains arbitrary in this construction.

In the case of $N=2$ QSCA, the non-linearity actually disappears  and the
algebra becomes the $N=2$ linear SCA, since the total coefficient in front of
the sum of two last terms in the second line of eq.~(2.4) vanishes ($i^2+1=0$)
after substituting $U(1)\cong SO(2)$ and the last eq.~(2.6). Therefore, the
non-linear structure of Bershadsky-Knizhnik QSCAs is only relevant when
$N\geq 3$.

The regular series of SCAs \cite{aba,aba2} with $N>4$ seem to be irrelevant
for constructing $N$-extended string theories~\footnote{The {\it unitary}
series of non-linear Bershadsky-Knizhnik QSCAs with $N>2$ are also irrelevant
in the search for new string theories because of another reason, see sect.~6.}
since they do {\it not} admit central extensions and always have subcanonical
charges \cite{aba,rs,htt}. In addition, there is no obvious $N$-extended
supersymmetric matter to represent world-sheet string fields in the case of
(Q)SCAs with $N>4$.~\footnote{Still, one may imagine that {\it all} string
world-sheet fields could be in one (presumably, non-linearly realised)
representation of $N>4$ QSCA. This would mean, however, very radical changes
in the philosophy underlying string theory nowadays. 2d supergravities based
on the QSCAs with $N>4$ may also exist.} In what follows, we are going to
consider the $N=4$ QSCAs in more details.

Eq.~(2.4) at $N=4$ takes the form
$$\eqalign{
G^i(z)G^j(w) ~\sim~ &~~ \fracmm{2k^2}{(k+1)}\fracmm{\d^{ij}}{(z-w)^3}
+\fracmm{2k}{(k+1)}\fracmm{(t^a)^{ij}J^a(w)}{(z-w)^2} ~,\cr
& +\fracmm{1}{z-w}\left[ 2\d^{ij}T(w)
+ \fracmm{k}{k+1}(t^a)^{ij}\pa J^a(w)\right]\cr
& +\fracmm{1}{(k+1)}\fracmm{1}{(z-w)}\left[ \left(t^{(a}t^{b)}\right)^{ij}
+ \d^{ab}\d^{ij}\right]:J^aJ^b:(w)~,\cr}\eqno(2.8)$$
whereas the Virasoro central charge of the $SO(4)$-based QSCA is simply
$$ c= 3k~.\eqno(2.9)$$

The $N=3$ Bershadsky-Knizhnik QSCA with $\Hat{SO(3)}$ KM symmetry can be
easily obtained by truncation of the $N=4$ algebra.

An analogue of the Sugawara-Sommerfeld relation for the $SO(4)$
Bershadsky-Knizhnik QSCA takes the weaker form \cite{sch2}
$$\pa T(z) = \fracmm{1}{4}:G^iG^i:(z) -\fracmm{1}{4(k+1)}\pa(:J^aJ^a:)(z)~,
\eqno(2.10)$$
which is consistent with all the commutation relations.
\vglue.2in

\section{$\hat{D}(1,2;\a)$ QSCA and `large' $N=4$ SCA}

The construction of the previous section in the case of $N=4$ QSCA is,
however, incomplete. Because of the isomorphism
$SO(4)\cong SU(2)\otimes SU(2)$, the supersymmetry generators transform in a
reducible $(2,2)$ representation of $SU(2)\otimes SU(2)$, while each of the
$\Hat{su(2)}$ KM components can have its own `level', thus opening the way for
generalisations. Therefore, the derivation of the $SO(4)$-based QSCA in
sect.~2 needs to be reconsidered, partly because of some additional identities
hold for $SO(4)$.

Let $J^{a\pm}(z)$ be the internal symmetry currents, where $a,b,\ldots$ are
now the adjoint indices of $SU(2)$, and $\pm$ distinguishes betweeen the two
$SU(2)$ factors. We still label the four-dimensional fundamental (vector)
representation space of $SO(4)$ by indices $i,j,\ldots\,$, as before. The
self-dual components of the KM currents, $J^{\pm a}(z)$ can be unified into an
antisymmetric tensor $J^{ij}(z)$ in the adjoint of $SO(4)$,
$$J^{ij}(z)=(t^{a-})^{ij}J^{a-}(z) +(t^{a+})^{ij}J^{a+}(z)~,\eqno(3.1)$$
where the antisymmetric $4\times 4$ matrices $t^{a\pm}$ satisfy the relations
$$\[t^{a\pm},t^{b\pm}\]=-2\ve^{abc}t^{c\pm}~,\quad \[t^{a+},t^{a-}\]=0~,\quad
\{t^{a\pm},t^{b\pm}\}=-2\d^{ab}~.\eqno(3.2)$$
These matrices can be explicitly represented as
$$(t^{a\pm})^{ij}=\ve^{aij}\pm (\d^i_a\d^j_4-\d^j_a\d^i_4)~,\eqno(3.3)$$
 and satisfy the identity
$$\sum_a (t^{a\pm})^{ij}(t^{a\pm})^{kl}=\d^{ik}\d^{jl}-\d^{il}\d^{jk}\pm
\ve^{ijkl}~.\eqno(3.4)$$
The OPEs describing the action of $J^{a\pm}(z)$ read
$$\eqalign{
J^{a\pm}(z)J^{b\pm}(w)~\sim~ & \fracmm{\ve^{abc}J^{c\pm}(w)}{z-w} +
\fracmm{-k^{\pm}\d^{ab}}{2(z-w)^2}~,\cr
J^{a\pm}(z)G^i(w)~\sim~ & \fracmm{{\frac 12}(t^{a\pm})^{ij}G^j(w)}{z-w}~,\cr}
\eqno(3.5)$$
where {\it two} arbitrary `levels' $k^{\pm}$ for both independent $\Hat{su(2)}$
 KM components have been introduced.

The general ansatz for the OPE of two fermionic supercurrents can be written as
$$
G^i(z)G^j(w)~\sim~  b_1 \fracmm{\d^{ij}}{(z-w)^3} +\fracmm{2T(w)\d^{ij}}{z-w}
 +\ha (b_2 + b_3) \left[\fracmm{J^{ij}(w)}{(z-w)^2} +\fracmm{{\frac 12}\pa
J^{ij}(w)}{z-w}\right]$$
$$ + \ha (b_2- b_3) \ve^{ijkl}\left[\fracmm{J^{kl}(w)}{(z-w)^2} +
\fracmm{{\frac 12}\pa J^{kl}(w)}{z-w}\right]
 +\fracmm{1}{4}b_4 \,\ve^{iklm}\ve^{jkpq}
\fracmm{:J^{lm}J^{pq}:(w)}{z-w}~, \eqno(3.6a)$$
in the vector notation, or, equivalently,
$$\eqalign{
G^i(z)G^j(w)~\sim~ & \fracmm{b_1\d^{ij}}{(z-w)^3} +\fracmm{1}{(z-w)^2}\left[
b_2(t^{a+})^{ij}J^{a+}(w) +b_3(t^{a-})^{ij}J^{a-}(w)\right] \cr
& \fracmm{1}{z-w}\left[ 2T(w)\d^{ij} + \ha\pa\left\{b_2(t^{a+})^{ij}J^{a+}(w)
 +b_3(t^{a-})^{ij}J^{a-}(w)\right\}\right] \cr
& +\fracmm{b_4}{z-w}:\left(t^{a+}J^{a+} -t^{a-}J^{a-}\right)^{(i}{}_k
\left(t^{b+}J^{b+} -t^{b-}J^{b-}\right)^{j)k}:(w)~,\cr}\eqno(3.6b)$$
in our new notation, where we have used the fact that
$$\ha\ve^{ijkl}\,T^{kl}(z)=\left(t^{a+}\right)^{ij}J^{a+}(z)
-\left(t^{a-}\right)^{ij}J^{a-}(z)~,\eqno(3.7)$$
as a consequence of eq.~(3.1). Compared to eq.~(2.4), note the presence of
extra terms in eq.~(3.6a) due to the $\ve$-symbol, and in eq.~(3.6b) due to
the $(\pm)$-distinction.

Demanding associativity of the combinations $TGG$, $JGG$ and $GGG$ determines
the parameters $b_1,\,b_2,\,b_3,\,b_4$, and, hence, all of the QSCA
3- and 4-point `structure constants', {\it viz.}
$$\eqalign{
b_1= \fracmm{4k^+k^-}{k^+ + k^- +2}~,\quad & \quad
b_4= \fracmm{-2}{k^+ + k^- +2}~,\cr
b_2= \fracmm{-4k^-}{k^+ + k^- +2}~,\quad & \quad
b_3= \fracmm{-4k^+}{k^+ + k^- +2}~,\cr}\eqno(3.8)$$
as well as the central charge,
$$c=\fracmm{6(k^++1)(k^-+1)}{k^++k^-+2}-3~,\eqno(3.9)$$
in agreement with refs.~\cite{bo,gs,gptvp}.

We define $\a$-parameter of this $\hat{D}(1,2;\a)$ QSCA as a ratio of its two
KM `levels',
$$\a\equiv  \fracmm{k^-}{k^+}~,\eqno(3.10)$$
which measures the relative asymmetry between the two $\Hat{su(2)}$ KM
algebras in the whole algebra. When $\a=1$, i.e. $k^-=k^+\equiv k$, the
$\hat{D}(1,2;1)$ QSCA is just the $SO(4)$ Bershadsky-Knihznik QSCA considered
in the previous section, with the central charge $c=3k$, as in eq.~(2.9).

In the vector notation, the $\hat{D}(1,2;\a)$ QSCA non-trivial OPEs take
the form
$$
T^{ij}(z)G^k(w)~\sim~ \fracmm{1}{z-w}\left[\d^{ik}G^{j}(w)-\d^{jk}G^{i}(w)
\right]~,\eqno(3.11a)$$
$$\eqalign{
J^{ij}(z)J^{kl}(w)~\sim~& \fracmm{1}{z-w}\left[\d^{ik}J^{jl}(w)-
\d^{jk}J^{il}(w)+\d^{jl}J^{ik}(w)-\d^{il}J^{jk}(w)\right]\cr
& - \ha(k^+ + k^-)\fracmm{\d^{ik}\d^{jl}-\d^{il}\d^{jk}}{(z-w)^2}
 - \ha(k^+ - k^-)\fracmm{\ve^{ijkl}}{(z-w)^2}~,\cr}\eqno(3.11b)$$
$$\eqalign{
G^i(z)G^j(w)~\sim~~~ &~\fracmm{4k^+k^-}{(k^+ + k^-+2)}\fracmm{\d^{ij}}{(z-w)^3}
+\fracmm{2T(w)\d^{ij}}{z-w}\cr
&  - \fracmm{k^+ + k^-}{k^+ + k^- +2}
\left[\fracmm{2J^{ij}(w)}{(z-w)^2} +\fracmm{\pa J^{ij}(w)}{z-w}\right]\cr
& + \fracmm{k^+ - k^-}{k^+ + k^- +2}\ve^{ijkl}
\left[\fracmm{2J^{kl}(w)}{(z-w)^2} +
\fracmm{\pa J^{kl}(w)}{z-w}\right] \cr
& - \fracmm{\ve^{iklm}\ve^{jkpq}}{2(k^+ + k^- +2)}
\fracmm{:J^{lm}J^{pq}:(w)}{(z-w)}~.\cr} \eqno(3.11c)$$
When using mode decompositions, like in eq.~(2.1), we find instead
$$\eqalign{
\[L_m,L_n\]= ~&~ (m-n)L_{m+n}+\fracmm{c}{12}m(m^2-1)\d_{m+n}~,\cr
\[L_m,G^i_r\]= ~&~ ({\frac 12}m-r)G^i_{m+r}~,\qquad
\[L_m,J^{a\pm}_n\]= -nJ^{a\pm}_m~,\cr
\[J^{a\pm}_m,G^i_r\]= ~&~ \ha(t^{a\pm})^{ij}G^j_{m+r}~,\qquad
\[J^{a\pm}_m,J^{b\pm}_n\]=  \ve^{abc}J^{c\pm}_{m+n} -{\frac 12}k^{\pm}m
\d^{ab}\d_{m+n}~,\cr
\{G^i_r,G^j_s\}= ~&~~~ \fracmm{2k^+k^-}{k^++k^-+2}\left(r^2-\ha\right)\d^{ij}
+2\d^{ij}L_{m+n}\cr
& +\fracmm{2}{k^++k^-+2}(s-r)\left[k^-(t^{a+})^{ij}J^{a+}_{r+s}
 + k^+(t^{a-})^{ij}J^{a-}_{r+s}\right]\cr
& -\fracmm{2}{k^++k^-+2}\left(t^{a+}J^{a+} - t^{a-}J^{a-}\right)^{(i}{}_k
\left(t^{b+}J^{b+} -t^{b-}J^{b-}\right)^{j)k}_{r+s}~.\cr}
\eqno(3.12)$$

Though $\hat{D}(1,2;\a)$ is a non-linear QSCA, it can be turned into a
{\it linear}
SCA by adding some `auxiliary' fields, namely, four free fermions $\j^i(z)$ of
dimension $1/2$, and a free Fubini-Veneziano bosonic current $U(z)$ of
dimension $1$, defining a $\Hat{U(1)}$ KM algebra \cite{gs}. The new fields
have canonical OPEs,
$$\eqalign{
\j^i(z)\j^j(w)~\sim~ & \fracmm{-\d^{ij}}{z-w}~,\cr
U(z)U(w)~\sim~ & \fracmm{-1}{(z-w)^2}~.\cr}\eqno(3.13)$$
The fermionic fields $\j^i(z)$ transform in a $(2,2)$ representation of
$SU(2)\otimes SU(2)$,
$$J^{a\pm}(z)\j^i(w)~\sim~ \fracmm{{\frac 12}(t^{a\pm})^{ij}\j^j(w)}{z-w}~,
\eqno(3.14)$$
whereas the singlet $U(1)$-current $U(z)$ can be thought of as derivative of
a free scalar boson, $U(z)=i\pa\f(z)$.

Let us now define the new currents \cite{gs}
$$\eqalign{
T_{\rm tot} = & T -\ha :U^2: - \ha:\pa\j^i\j^i:~,\cr
G^i_{\rm tot} = & G^i - U\j^i
+\fracmm{1}{3\sqrt{2(k^+ +k^-+2)}}\ve^{ijkl}\j^j\j^k\j^l\cr
&  -\sqrt{\fracmm{2}{k^+ +k^-+2}}\,\j^j\left[(t^{a+})^{ji}J^{a+}-
(t^{a-})^{ji}J^{a-}\right]~,\cr
J^{a\pm}_{\rm tot} = & J^{a\pm} +\fracmm{1}{4}(t^{a+})^{ij}\j^i\j^j~,\cr}
\eqno(3.15)$$
in terms of the initial $\hat{D}(1,2;\a)$ QSCA currents
$T~,\,G^i$ and $J^{a\pm}$. Then the following set of affine generators
$$\left\{ T_{\rm tot}~,\quad  G^i_{\rm tot}~,\quad J^{a\pm}_{\rm tot}~,\quad
\j^i~,\quad U\,\right\}\eqno(3.16)$$
has closed OPEs among themselves, defining a linear `large' $N=4$ SCA
with the $\Hat{su(2)}\oplus\Hat{su(2)}\oplus\Hat{u(1)}$ KM component!
Explicitly, the non-trivial OPEs of this `large' $N=4$ SCA are given by
({\it cf} refs.~\cite{sch1,stvp,gptvp})
$$\eqalign{
T_{\rm tot}(z)T_{\rm tot}(w)~\sim~ & \fracmm{{\frac 12}(c+3)}{(z-w)^4}
 + \fracmm{2T_{\rm tot}(w)}{(z-w)^2} + \fracmm{\pa T_{\rm tot}(w)}{z-w}~,\cr
T_{\rm tot}(z)\co(w)~\sim~ & \fracmm{h_{\co}\co(w)}{(z-w)^2} +
 \fracmm{\pa \co (w)}{z-w}~,\cr
J_{\rm tot}^{a\pm}(z)J_{\rm tot}^{a\pm}(w)~\sim~ &
\fracmm{\ve^{abc}J_{\rm tot}^{c\pm}(w)}{z-w} - \fracmm{(k^{\pm}+1)\d^{ab}}{2
(z-w)^2}~,\cr
J_{\rm tot}^{a\pm}(z)G^i_{\rm tot}(w)~\sim~ &
\fracmm{{\frac 12}(t^{a\pm})^{ij}G^j_{\rm tot}(w)}{z-w}  \mp
\fracmm{k^{\pm}+1}{\sqrt{2(k^++k^-+2)}}
\fracmm{(t^{a\pm})^{ij}\j^j(w)}{(z-w)^2}~,\cr
G^i_{\rm tot}(z)G^j_{\rm tot}(w)~\sim~ &
\fracmm{{\frac 23}(c+3)\d^{ij}}{(z-w)^3} +\fracmm{2T_{\rm tot}(w)\d^{ij}}{z-w}
 -\fracmm{2}{k^++k^-+2}\left[\fracmm{2}{(z-w)^2}+\fracmm{1}{z-w}\pa_w\right]\cr
& \times \left[(k^-+1)(t^{a+})^{ij}J^{a+}_{\rm tot}(w)+
(k^++1)(t^{a-})^{ij}J^{a-}_{\rm tot}(w)\right]~,\cr
\j^i(z)G^j_{\rm tot}(w)~\sim~ & \fracmm{1}{z-w}\sqrt{
\fracmm{2}{k^++k^-+2}}\left[ (t^{a+})^{ij}J^{a+}_{\rm tot}(w)
-(t^{a-})^{ij}J^{a-}_{\rm tot}(w)\right] +\fracmm{U(w)\d^{ij}}{z-w}~,\cr
U(z)G_{\rm tot}^i(w)~\sim~ & \fracmm{\j^i(w)}{(z-w)^2}~,\cr}\eqno(3.17)$$
where $\co$ stands for the generators $G_{\rm tot},\,J_{\rm tot}$ and $\j$ of
dimension $3/2,\,1$ and $1/2$, respectively, and the $\hat{D}(1,2;\a)$ QSCA
central
charge $c$ is given by eq.~(3.9). Unlike refs.~\cite{stvp,gptvp}, we always
put forward the underlying QSCA structure in our notation.

Having restricted ourselves to the (Neveu-Schwarz--type, for definiteness)
modes~\footnote{ See eq.~(2.1) for their definition.}
$(L_{\rm tot})_{\pm 1,0}~,~~(G^i_{\rm tot})_{\pm 1/2}$ and
$(J_{\rm tot}^{a\pm})_0\,$, we get a finite-dimensional Lie
superalgebra which is isomorphic to the simple Lie superalgebra $D(1,2;\a)$
from the Ka\v{c} list \cite{kac}. This explains the reason why we use almost
the same (with hat) notation for our affine (infinite-dimensional) QSCA
$\hat{D}(1,2;\a)$ defined by eqs.~(3.11) or (3.12). Note that the finite Lie
superalgebra of the `large' $N=4$ SCA in eq.~(3.17), defining a `linearised'
version of the $\hat{D}(1,2;\a)$ QSCA in eq.~(3.11), is not simple, but
contains a $U(1)$ piece, in addition to the finite-dimensinal $D(1,2;\a)$
subalgebra. The finite-dimensional simple Lie superalgebras $D(2,1;\a)$ at
various $\a$ values are not, in general, isomorphic to each other (except of
the isomorphism under $\a\to\a^{-1}$, interchanging the two $su(2)$ factors)
\cite{kac}. This is enough to argue about the non-equivalence
(for different $\a$) of the $\hat{D}(1,2;\a)$ QSCAs, which are their affine
generalisations.

It is also worthy to notice that the KM `levels' and the central charge of the
`large' $N=4$ SCA and those of the underlying $\hat{D}(1,2;\a)$ QSCA are
different according to eq.~(3.16), namely
$$ k^{\pm}_{\rm large} = k^{\pm}+1~,\qquad c_{\rm large} = c+3~,\eqno(3.18)$$
which is quite obvious because of the new fields introduced. The exceptional
`small' $N=4$ SCA with the $\Hat{su(2)}$ KM component \cite{aba} follows from
eq.~(3.17) in the limit $\a\to\infty$ or $\a\to 0$, where either $k^-\to\infty$
or $k^+\to\infty$, respectively, and the $\Hat{su(2)}\oplus\Hat{u(1)}$ KM
component decouples from the rest of the algebra. Taking the limit results in
the central charge
$$ c_{\rm small}=6k~,\eqno(3.19)$$
where $k$ is an arbitrary `level' of the remaining $\Hat{su(2)}$ KM component.
For an arbitrary $\a$, the `large' $N=4$ SCA contains two `small' $N=4$
SCAs~\cite{stvp}.

As regards the representation theory, it is also advantageous to express a
given algebra in terms of the smaller number of fundamental charges, whenever
it is possible, since it makes more evident the structure of its
representations. The representation theory of the $SO(N)$-based
Bershadsky-Knizhnik QSCAs was developed in ref.~\cite{sch2}. Some of these
representations can be naturally constructed by using the Kazama-Suzuki method
on quaternionic (Wolf) spaces \cite{vp}. A construction of unitary highest
weight (positive energy) representations of the `small' and `large' $N=4$ SCAs
can be found in ref.~\cite{et} and refs.~\cite{sstvp,gptvp}, respectively.
Since both the $\hat{D}(1,2;\a)$ QSCA and the `large' $N=4$ SCA contain two
$\Hat{SU(2)}$ KM subalgebras, unitary requires both levels $k^{\pm}$ to be
non-negative integers, in particular. The Kazama-Suzuki method on the
quaternionic spaces ${\it Wolf}\otimes SU(2)\otimes U(1)$ can also be used to
get unitary highest weight representations of the `large' $N=4$ SCA for
$3\leq c_{\rm large}<6\,$ \cite{sstvp}.

The `basic' unitary (non-linear) representation \cite{sch1} of the
`large' $N=4$ SCA in terms of four free fermions and one boson $(\j^i,\f)$ can
be immediately obtained from eq.~(3.15) by taking $T(z)=G^i(z)=J^{a\pm}(z)=
0$. This representation has the lowest possible central charge value, $c=3$,
among all the unitary representations of the `large' $N=4$ algebra, having the
central charge (2.9) for non-negative integers $k$. This representation is
actually realised in a model with defected Ising chains \cite{hp}. More
unitary representations can be constructed by taking tensor products of the
`basic' representation \cite{sch1,stvp}.
\vglue.2in

\section{Gauging a Classical QSCA}

2d quantum conformal field theory requires the closure of its OPEs \cite{book}.
There is nothing wrong there if the closure actually takes place in the
 $W$-sense, i.e. in terms of (regularised) local products of the
fundamental fields. When building a $W$-type (super)gravity or a $W$-type
(super)string, we need to represent the underlying non-linear algebra as a
symmetry algebra, with the symmetries to be locally realised.

All affine algebras considered in the previous sections are the quantum
(Q)SCAs. Their classical versions can be recovered in the limit $\hbar\to 0$,
after making the substitutions $\co\to\tilde{\co}\equiv\hbar\co$ for all the
(Q)SCA generators $\co$, and using the standard correspondence between the
classical (graded) Poisson brackets and the (anti)commutators,
$$\{ \tilde{\co}_1,\tilde{\co}_2 \}_{\rm P.B.}=\lim_{\hbar\to 0}
\fracmm{1}{\hbar}\[\tilde{\co}_1,\tilde{\co}_2 \]~.\eqno(4.1)$$
It should be stressed, however, that, in order to get a correct classical
result, one should take into account an additional factor of $\hbar$ in front
of the current bilinear in a quantum QSCA.

Therefore, when starting from a quantum QSCA, we first need to identify its
{\it classical} analogue and its {\it non-anomalous} component. Both are of
crucial importance for the gauging procedure. The main goal of this section is
 to outline a construction of the new ($W$-type) $N=4$ extended conformal
supergravitiy based on the classical version of the non-linear quantum algebra
$\hat{D}(1,2;\a)$ introduced in the previous section. We are going to compare
this new $N=4$ conformal supergravity with the standard $SU(2)$-based $N=4$
conformal supergravity in two dimensions. More details about the new $N=4$
supergravity are going to be reported in a separate publication \cite{gaket}.

As to the regular (linear) $N$-extended SCAs constructed by Ademollo et
al.~\cite{aba}, the corresponding classical symmetry algebras
with vanishing central terms can be easily realised in terms of the restricted
 (=~superconformal) superdiffeomorphisms in the $N$-extended light-cone
superspace \cite{aba}, where the vanishing $N$-extended schwartzian derivative
just means no anomaly. The last condition is known to be solved by the
superprojective transformations forming the  `little superconformal
group' $OSP(N/2)$ \cite{sch3}. The regular SCAs do {\it not} admit central
extensions for $N\geq 5$ \cite{sch4}, and, therefore, they are of no interest
for superconformal quantum field theory where anomalous terms in the
transformation laws correspond to central terms in the current algebra. The
case of $N=4$ is special: though the `large' $N=4$ SCA does admit central
extensions, it cannot be gauged in full because of the presence of matter
$(\j^i,\f)$ among its `currents'. The only case of linear quantum $N=4$ SCA
which can be locally realised is the exceptional `small' $N=4$ SCA equivalent
to the $SU(2)$-based Bershadsky-Knizhnik (Q)SCA \cite{kn,be}. The corresponding
2d, $N=4$ conformal supergravity was constructed by gauging this `small' $N=4$
 SCA in ref.~\cite{sch3}. The $N=4$ locally supersymmetric string
action was known even before \cite{pn}.

As far as the non-linear $\hat{D}(1,2;\a)$ QSCA is concerned, its classical
limit $(~\equiv~\tilde{D}_4)$ with vanishing central terms takes the form
$$\eqalign{
\{T(\z^+),T(\x^+)\}_{\rm P.B.} = ~&~ -\d'(\z^+ - \x^+)\left[T(\z^+)
+ T(\x^+)\right]~,\cr
\{T(\z^+),G^i(\x^+)\}_{\rm P.B.} = ~&~ -\d'(\z^+ - \x^+)\left[G^i(\z^+) +
\ha G^i(\x^+)\right]~,\cr
\{T(\z^+),J^{a\pm}(\x^+)\}_{\rm P.B.} = ~&~ -\d'(\z^+ -\x^+)\left[J^{a\pm}
(\z^+) -\ha J^{a\pm}(\x^+)\right]~,\cr
\{J^{a\pm}(\z^+),G^i(\x^+)\}_{\rm P.B.} = ~&~ \d(\z^+ - \x^+)\ha
(t^{a\pm})^{ij}G^j(\x^+)~,\cr
\{J^{a\pm}(\z^+),J^{b\pm}(\x^+)\}_{\rm P.B.} = ~&~ \d(\z^+ - \x^+)\,\ve^{abc}
J^{c\pm}(\x^+)~,\cr
\{G^i(\z^+),G^j(\x^+)\}_{\rm P.B.} = ~&~ \d(\z^+ -\x^+)\left[2\d^{ij}T(\x^+)
- \L^{ij}(\x^+)\right]~,\cr}\eqno(4.2)$$
where the new composite generator $(A=+,-)$
$$\eqalign{
\L(\z^+)^{ij} & = \L_{aAbB}^{ij}J^{aA}(\z^+)J^{bB}(\z^+)\cr
& \equiv  \left(t^{a+}J^{a+}(\z^+)-t^{a-}J^{a-}(\z^+)
\right)^{(i}{}_k \left(t^{b+}J^{b+}(\z^+) -t^{b-}J^{b-}(\z^+)
\right)^{j)k}~,\cr}\eqno(4.3)$$
and the 2d space-time light-cone coordinates $\z^{\pm}$, related
to the usual 2d Cartesian coordinates $(t,\s)$ by
$\z^{\pm}=\fracmm{1}{\sqrt{2}}\left(t\pm\s\right)$,
have been introduced. The holomorphic coordinate $z$, appearing in the OPEs,
is related to them via the Wick rotation, $\t=it$,
and the exponential map, $z=e^{\t+i\s}$, as is usual in conformal field theory
\cite{book}. We also have to assume here that all currents in eq.~(4.2) arise
from some classical conformal field theory, so that the bracket on the l.h.s.
of that equation is the graded Poisson bracket in a canonical formalism with
$\z^-$ as the time variable. Note that, in classical theory, there is no
problem in defining the product of currents in eq.~(4.3) at coincidence limit,
unlike the situation in quantum theory where one uses normal ordering and
divergence subtraction. The classical algebra $\tilde{D}_4$ defined by
eqs.~(4.2) and (4.3) can be considered as a particular supersymmetric version
of the Gel'fand-Dikii-type algebras known in the theory of integrable models
\cite{gd}.

When following the lines of construction of the conventional extended
conformal supergravity theories in two dimensions \cite{vh}, it would be
natural to pick up
$$  L_{\pm 1,0}~,\qquad G^i_{\pm 1/2}~,\qquad J^{a\pm}_0~, \eqno(4.4)$$
as the non-anomalous generators. However, according to eq.~(3.12), they do {\it
not} form a closed subalgebra in the case of $\hat{D}(1,2;\a)$, because of the
non-linear term present in the anticommutator of the supersymmetry charges,
$$\eqalign{
\{ G^i_r,G^j_s\} = & 2\d^{ij}L_{r+s}
+\fracmm{2}{k^++k^-+2}(s-r)\left[k^-(t^{a+})^{ij}J^{a+}_{r+s}
 + k^+(t^{a-})^{ij}J^{a-}_{r+s}\right]\cr
& -\fracmm{2}{k^++k^-+2}\left(t^{a+}J^{a+} - t^{a-}J^{a-}\right)^{(i}{}_k
\left(t^{b+}J^{b+} -t^{b-}J^{b-}\right)^{j)k}_{r+s}~,\cr} \eqno(4.5)$$
where $r,s=\pm1/2,\quad r+s=-1,0,1$. Though the second term on the
r.h.s. of this equation never contributes when $s=r=\pm{\frac 12}$ and, hence,
it does not actually depend on $J^{a\pm}_{\pm 1}$, the last term
vanishes only in the limit $k^++k^-\equiv 2k\to\infty$ where
the finite set (4.4) formally constitutes a linear algebra.~\footnote{This is
to be compared with the known fact that, e.g. in the Zamolodchikov's $W_3$
algebra, the coefficient in front of the only non-linear term in the commutator
 of two spin-$3$ generators vanishes in the limit $c\to\infty$, where $c$ is
the quantum $W_3$ central charge \cite{za}.}

The `small' (non-chiral) $SU(2)$-based $N=4$ conformal supergravity \cite{pn}
is known to be obtained by gauging the linear $ssu(1,1|2)\oplus ssu(1,1|2)$
Lie superalgebra, whereas its chiral $(4,0)$ supersymmetric version is based on
a factor $ssu(1,1|2)$ \cite{sch3}. The superalgebra $ssu(1,1|2)$ appears, in
particular, in the Ka\v{c} list of {\it finite}-dimensional simple Lie
superalgebras \cite{kac}, and its internal symmetry generators are just the
self-dual or anti-self-dual ones, $J^{a+}$ {\it or} $J^{a-}$. When comparing
its contents and (anti)commutation relations with what we have in eq.~(4.2),
it is clear that conformal supergravity based on gauging the classical
$W$-type non-linear algebra $\tilde{D}_4$ is going to be different from the
 `small' $N=4$ conformal supergravity. Indeed, the former has six (instead of
three) spin-$1$ gauge fields in the vector representation of $SO(4)$, and its
underlying algebra is non-linear.

The full 2d algebra to be gauged is the direct sum of two light-cone parts,
each one being isomorphic to $\tilde{D}_4$. Gauging only one factor should
correspond to a `chiral' $\tilde{D}_4$ supergravity. It seems to be quite
appropriate to identify it as the classical $(4,0)$ supersymmetric `heterotic'
(or chiral) $\tilde{D}_4$ conformal supergravity, whereas the full (non-chiral)
2d theory based on gauging the $\tilde{D}_4\oplus\tilde{D}_4$ classical algebra
should be called the $(4,4)$ supersymmetric (non-chiral) $\tilde{D}_4$
conformal supergravity. Because of the classical isomorphisms
$su(1,1)\sim so(2,1)\sim sl(2)$ and $so(2,2)\sim sl(2)\oplus sl(2)$, the full
algebra $\tilde{D}_4\oplus\tilde{D}_4$ obviously contains the 2d `little'
(finite-dimensional) conformal algebra $so(2,2)$, as it should \cite{nrev}.
In addition, it has four supersymmetry charges of each chirality. Hence, we
are dealing with an $N=4$ extended conformal 2d supergravity indeed.

Consider now any 2d classical superconformal field theory with the
$\tilde{D}_4$ or $\tilde{D}_4\oplus\tilde{D}_4$ symmetry. It could be, e.g., a
$(4,0)$ or $(4,4)$ supersymmetric WZNW model on a quaternionic (Wolf)
space.~\footnote{The Wolf spaces are also selected for the couplings of $N=2$
scalar multiplets to $N=2$ extended supergravity in {\it four} dimensions
\cite{bw}.} Let $S_0$ be its classical action, and $W_{\pm M}$ the
$\tilde{D}_4\oplus\tilde{D}_4$ currents labelled by some (generalised) index
$M$ and satisfying the constraints $\pa_{\mp}W_{\pm M}=0$ where the signs are
correlated. Then the gauging procedure is very similar to the one known for
the classical  $W$ algebras.~\footnote{See, e.g., refs.~\cite{hrev,svp}
for a review.} One introduces gauge fields $h^{\pm M}$ for each current and
adds the Noether (minimal) coupling to $S_0$. To lowest order, the
action is given by \cite{hull}
$$S = S_0 + \sum_M \int d^2\z\,\left[ h^{+M}W_{+M} + h^{-M}W_{-M}\right]
+O(h^2)~,\eqno(4.6)$$
where the higher order corrections could, in principle, be calculated using
the Noether (trial and error) method. Transformations laws of the gauge
fields get fixed by imposing the (anti)commutation relations of the symmetry
algebra $\tilde{D}_4\oplus\tilde{D}_4$. As is usual for the classical
$W$ algebras, the Noether coupling alone gives the full gauge-invariant action
for the {\it chiral} gauge theory of $\tilde{D}_4$, where only the gauge
fields $h^{+M}$ are present \cite{hull}.

In the non-chiral case, the full gauge-invariant action is non-polynomial in
the gauge fields, and, in a non-covariant form, can be constructed within
the canonical approach as follows \cite{mic}.~\footnote{The related approach
based on introducing auxiliary fields \cite{ssn} can also be used for this
purpose.} After rewriting the action
$S_0$ to the hamiltonian (first-order) form and replacing time derivatives of
fields in the $\tilde{D}_4$ currents by the corresponding momenta, the
gauge-invarint action is obtained by simply adding the Noether coupling of
these currents to Lagrange multipier gauge fields. The full action reads
$$S = \int dt\,\left[p_A\pa_tq^A -h^MW_{M}(p,q)\right]~,\eqno(4.7)$$
in terms of the generalised coordinates $q^A(t)$ and the momenta
$p_A(t)$.~\footnote{We use here the condensed notation in which the
generalised indices $A,B,\ldots$ represent both the discrete indices and the
continuous variable $\s$. The same convention applies to summations.} The
gauge fields impose the first-class constraints $W_M\sim 0$, whose Poisson
bracket algebra obviously closes in the weak sense (on the constraints),
$$ \{W_M,W_N\}_{\rm P.B.}=f_{MN}{}^L(p,q)W_L~,\eqno(4.8)$$
where some of the structure `constants' $f_{MN}{}^L$ in the case of the
non-linear algebra $\tilde{D}_4\oplus\tilde{D}_4$  are {\it dependent} on phase
space variables. The action (4.8) is invariant under the following local
symmetries with parameters $\ve^M(t)$:
$$\eqalign{
\d p_A=~&~\ve^M\{W_M,p_A\}_{\rm P.B.}~,\cr
\d q^A=~&~\ve^M\{W_M,q^A\}_{\rm P.B.}~,\cr
\d h^M=~&~\pa_t\ve^M -f_{NL}{}^Mh^N\ve^L~.\cr}\eqno(4.9)$$
Elimination of momenta is, however, non-trivial in the action (4.7), but this
is a technical problem.

In summary, the gauge field contents of the 2d, $\tilde{D}_4$ conformal
supergravity is
$$ h_{\m\n}~,\qquad \c^i_{\m}~,\qquad A^{a\pm}_{\m}~,\eqno(4.10)$$
where $h_{\m\n}$ is spin-$2$ graviton, $\c^i_{\m}$ are four spin-$3/2$
Majorana gravitinos, and $A^{a\pm}_{\m}$ are six spin-$1$ gauge fields.
The gauge field contents is balanced by the gauge symmetries as usual, which
implies no off-shell degrees of freedom for this new $SO(4)$-extended 2d
conformal supergravity. In other words, the classical 2d, $\tilde{D}_4$
conformal supergravity is trivial (up to moduli). However, in quantum theory,
some of the gauge symmetries may become anomalous and thereby some of the
gauge degrees of freedom may become physical. The ghost contributions to
the central charge of the BRST quantised $N=4$ conformal supergravities are
collected in Table I (see sect.~5 for a derivation and more results).
\begin{center}
{\sf Table I}. The ghost contributions to the $N=4$ central charge.
\vglue.1in
\noindent\begin{tabular}{cccc} \hline
dimension & $c_j$ & `small' SCA & $\hat{D}(1,2;\a)$   \\
\hline
$2$ & $-26$ & $-26$ & $-26$ \\
$3/2$ & $+11$ & $+44$ & $+44$ \\
$1$ & $-2$ & $-6$ & $-12$ \\
\hline
& total & $+12$ & $+6$ \\
\hline
\end{tabular}
\vglue.2in
\end{center}

\section{BRST Charge}

In this section, the nilpotent quantum BRST charges for the non-linear QSCAs
introduced in sects.~2 and 3 are constructed. Despite of the apparent
non-linearity of these algebras, their quantum BRST charges should be in
correspondence with their classical BRST charges, up to renormalisation. A
classical BRST charge having the vanishing Poisson bracket with itself can, in
fact, be constructed for an arbitrary algebra of first-class
constraints~\cite{ff}.~\footnote{See Appendix for details.} This procedure
was already applied to obtain the quantum BRST charge for the
non-linear quantum $W_3$ algebra \cite{tm}, and later generalised to any
quadratically non-linear $W$-type algebra in ref.~\cite{ssvn}. In particular,
the quantum BRST charges for the orthogonal and unitary series of
Berschadsky-Knizhnik QSCAs were also calculated in ref.~\cite{ssvn}, but the
analysis of the BRST charge nilpotency conditions given there was, however,
incomplete. The nilpotency conditions always
require the total (matter + ghosts) central charge to vanish, but also lead to
some more constraints on the QSCA parameters, whose consistency is {\it not}
guaranteed.  This is because the `new' constraints imposed by the
BRST charge nilpotency condition may be in conflict with the `old' constraints
dictated by the QSCA Jacobi identities. Our calculations of the quantum BRST
charge for the Berschadsky-Knizhnik $SO(N)$-based QSCAs confirm `almost'
all of the results of ref.~\cite{ssvn}, but one. Namely, we find that the {\it
only} nilpotent solution exists at $N=4$. We then make a similar calculation
for the new case of the $N=4$ supersymmetric $\hat{D}(1,2;\a)$ QSCA, obtain
the corresponding quantum BRST charge and its nilpotency conditions. Again, we
find only one solution, namely, just in the case when the $\hat{D}(1,2;\a)$
QSCA reduces to the $SO(4)$ Berschadsky-Knizhnik QSCA. Some consequences of
our results for string theory are discussed in Conclusions, sect.~6.

We begin with the Berschadsky-Knizhnik $SO(N)$-based QSCA introduced in
sect.~2. The BRST ghosts appropriate for this case are:
\begin{itemize}
\item the conformal ghosts ($b,c$), an anticommuting pair of
world-sheet free fermions of conformal dimensions~($2,-1$), respectively;
\item the $N$-extended superconformal ghosts ($\b^i,\g^i$) of conformal
dimensions~($\frac32,-\frac12$), respectively, in the fundamental (vector)
representation of $SO(N)$;
\item the $SO(N)$ internal symmetry ghosts ($\tilde{b}^a,\tilde{c}^a$) of
conformal dimensions~($1,0$), respectively, in the adjoint representation of
$SO(N)$.
\end{itemize}

The reparametrisation ghosts
$$b(z)\ =\ \sum_{n\in{\bf Z}} b_n z^{-n-2}~,\qquad
c(z)\ =\ \sum_{n\in{\bf Z}} c_n z^{-n+1}~,\eqno(5.1)$$
have the following OPE and  anticommutation relations:
$$b(z)\ c(w)\ \sim\ \fracmm{1}{z-w}~,
\qquad \{c_m,b_n\}\ =\ \d_{m+n,0}~.\eqno(5.2)$$

The superconformal ghosts
$$\b^i(z)\ =\ \sum_{r\in{\bf Z}(+1/2)}\b^i_r z^{-r-3/2}~,\qquad
\g^i(z)\ =\ \sum_{r\in{\bf Z}(+1/2)}\g^i_r z^{-r+1/2}~,\eqno(5.3)$$
satisfy the OPE
$$\b^i(z)\ \g^j(w)\ \sim\ \fracmm{-\d^{ij}}{z-w}~, \eqno(5.4)$$
which implies the only non-vanishing commutation relations
$$\[\g^i_r,\b^j_s\]\ =\ \d_{r+s,0}~.\eqno(5.5)$$
An integer or half-integer moding of these generators corresponds to the usual
distinction between the Ramond- and Neveu-Schwarz--type sectors.

Finally, the fermionic $SO(N)$ ghosts
$$\tilde{b}^a(z)\ =\ \sum_{n\in{\bf Z}}\tilde{b}^a_n z^{-n-1}~,\qquad
\tilde{c}^a(z)\ =\ \sum_{n\in{\bf Z}} \tilde{c}^a_n z^{-n}~,\eqno(5.6)$$
have
$$\tilde{b}^a(z)\ \tilde{c}^a(w)\ \sim\ \fracmm{\d^{ab}}{z-w}~,\qquad
\{\tilde{c}^a_m,\tilde{b}^b_n\}\ =\ \d^{ab}\d_{m+n,0}~.\eqno(5.7)$$

The construction of a classical BRST charge for any (non-linear) algebra of
first-class constraints is reviewed in Appendix. This construction provides
 us with the reasonable ansatz for the {\it quantum} BRST charge~\footnote{The
normal ordering is implicit below.} associated with the quantum $SO(N)$-based
Bershadsky-Knizhnik QSCA in the form ({\it cf~} ref.~\cite{ssvn})
$$\eqalign{
Q_{\rm BRST}=~&~ c_{-n}L_n + \g^i_{-r}G^i_r + \tilde{c}^a_{-n}J^a_n
-\ha (m-n)c_{-m}c_{-n}b_{m+n} + nc_{-m}\tilde{c}^a_{-n}\tilde{b}^a_{m+n}\cr
& +\left(\fracm{m}{2}-r\right)c_{-m}\b^i_{m+r}\g^i_{-r}
-b_{r+s}\g^i_{-r}\g^i_{-s} - \tilde{c}^a_{-m}\b^i_{m+r}(t^a)^{ij}\g^j_{-r}\cr
& + \h a_2(r-s)\tilde{b}^a_{r+s}\g^i_{-r}(t^a)^{ij}\g^j_{-s}
-\ha f^{abc}\tilde{c}^a_{-m}\tilde{c}^b_{-n}\tilde{b}^c_{m+n} \cr
& -\fracmm{1}{2}a_4\left[\left(t^{(a}t^{b)}\right)^{ij}+\d^{ab}\d^{ij}\right]
J^a_{r+s+m}\tilde{b}^b_{-m}\g^i_{-r}\g^j_{-s}
 -\fracmm{1}{24}a^2_4\left[\left(t^{(a}t^{b)}\right)^{ij}+\d^{ab}\d^{ij}\right]
\cr
&\times \left[\left(t^{(c}t^{d)}\right)^{kl}+\d^{cd}\d^{kl}\right]
 f^{ace}\d_{m+n+p,r+s+t+u}\tilde{b}^b_m\tilde{b}^d_n\tilde{b}^e_p
 \g^i_{-r}\g^j_{-s}\g^k_{-t}\g^l_{-u}~,\cr}\eqno(5.8)$$
where a quantum renormalisation parameter $\h$ has been introduced. Its value
is going to be fixed by the BRST charge nilpotency conditions. The coefficients
$a_2$ and $a_4$ have already been fixed by eq.~(2.6).

We find always useful to represent a quantum BRST charge as
$$Q_{\rm BRST}=\oint_0 \fracmm{dz}{2\p i}\,j_{\rm BRST}(z)~,\eqno(5.9)$$
where the BRST current $j_{\rm BRST}(z)$ is defined {\it modulo} total
derivative.~\footnote{The total derivative can be fixed by requring the
$j_{\rm BRST}(z)$ to transform as a primary field.} In particular, the BRST
current $j_{\rm BRST}(z)$ corresponding to the BRST charge of eq.~(5.8) is
given by
$$\eqalign{
j_{\rm BRST}(z)=~&~ cT +\g^i G^i + \tilde{c}^a J^a +bc\pa c
-c\tilde{b}^a\pa\tilde{c}^a
 -\ha c\g^i\pa\b^i-\fracm{3}{2}c\b^i\pa\g^i
-b\g^i\g^i\cr
& -\h a_2
\tilde{b}^a (t^a)^{ij}\left(\g^i\pa\g^j-\g^j\pa\g^i\right)
-\tilde{c}^a(t^a)^{ij}\b^i\g^j
 -\ha f^{abc}\tilde{c}^a\tilde{c}^b\tilde{b}^c\cr
& -\ha a_4
\left[\left(t^{(a}t^{b)}\right)^{ij}+\d^{ab}\d^{ij}\right]
J^a\tilde{b}^b\g^i\g^j  -\fracmm{1}{24} a^2_4
\left[\left(t^{(a}t^{b)}\right)^{ij}+\d^{ab}\d^{ij}\right]\cr
& \times \left[\left(t^{(c}t^{d)}\right)^{kl}+\d^{cd}\d^{kl}\right]f^{ace}
\tilde{b}^b\tilde{b}^d\tilde{b}^e\g^i\g^j\g^k\g^l~.\cr}\eqno(5.10)$$

The quantum BRST charge for any (quasi)SCA can always be written in the
standard form,
$$ Q_{\rm BRST} = \oint\fracmm{dz}{2\p i}\,
\left[c(T+\ha T_{\rm gh})+\g^i(G^i + \ha G^i_{\rm gh})+ \tilde{c}^{a}(J^a +\ha
J^a_{\rm gh})\right]~,\eqno(5.11)$$
where the ghost contributions which can be read off from an explicit
formula for $Q_{\rm BRST}$. There are, however, some important  differences
between the linear and non-linear (Q)SCAs because, as is clear, e.g., from
eq.~(5.10), the QSCA ghost contributions are {\it dependent} on the matter
currents and include terms of higher order in the (anti)ghosts. In particular,
 the ghost
supercurrent $G^i_{\rm gh}$ in eqs.~(5.10) and (5.11) involves the spin-$1$
matter currents $J^a$, which is intuitively reasonable since one can view the
non-linear bilinear on the r.h.s. of the supersymmetry algebra in
eq.~(2.8) as being like a linear algebra but with $J$-dependent structure
`constants'. These structure constants then appear in the ghost currents. The
ghost currents alone need not satisfy the QSCA, and they actually do not.
In addition, central extensions (anomalies) of the ghost-extended QSCA need
not form a linear supermultiplet, and they actually do not also. Therefore, the
 vanishing of any anomaly alone does {\it not} automatically imply the
vanishing of the others, unlike in the linear case.

The most tedious part of calculational handwork in computing $Q^2_{\rm BRST}$
can be avoided when using either the Mathematica Package for computing OPEs
\cite{th} or some of the general results in ref.~\cite{ssvn}. In particular,
as was shown in ref.~\cite{ssvn}, quantum renormalisation of the $3$-point
structure constants in the quantum BRST charge should be {\it multiplicative},
whereas the non-linearity $4$-point `structure constants' should {\it not} be
renormalised at all --- the facts already used in the BRST charge ansatz above.
 Most importantly, among the contributions to the $Q_{\rm BRST}^2$, only the
terms {\it quadratic} in the ghosts are relevant. Their vanishing imposes
 the constraints on the central extension coefficients of the QSCA and
simultaneously determines the renormalisation parameter $\h$. The details
can be found in the appendices of ref.~\cite{ssvn}. The same conclusion comes
as a result of straightforward calculation on computer. Therefore, finding out
the nilpotency conditions amounts to calculating only a few terms `by hands',
namely, those which are quadratic in the ghosts. This makes the whole
calculation as simple as that in ordinary string theories based on linear SCAs
 \cite{book}.

The $2$-ghost terms in the $Q_{\rm BRST}^2$ arise from single contractions of
the first three linear (in the ghosts) terms of $Q_{\rm BRST}$ with themselves
and with the next cubic terms of eq.~(5.10), and from double contractions of
the latter among themselves. They result in the pole contributions to
$j_{\rm BRST}(z)j_{\rm BRST}(w)$, proportional to $(z-w)^{-n}$
with $n=1,2,3,4$. All the residues have to vanish modulo total derivative. We
find
$$\eqalign{
j_{\rm BRST}(z)j_{\rm BRST}(w)~\sim~&
 \fracmm{c(z)c(w)}{2(z-w)^4}\left[c-N^2+12N-26\right]  \cr
{}~&~+\fracmm{\g^i(z)\g^i(w)}{(z-w)^3}\left[a_1 -\fracmm{ka_4}{2}(N-1)(N-2)
-4\h a_2(N-1) +2\right]  \cr
{}~&~+\fracmm{\tilde{c}^a(z)\tilde{c}^a(w)}{(z-w)^2}\left[-k-2(N-2)+2\right]
\cr
{}~&~+\fracmm{J^a(w)(t^a)^{ij}\g^i(w)\pa\g^j(w)}{z-w}\left[-4\h a_2
-4a_4\left(1-\fracmm{N}{2}\right)\right]+\ldots~,  \cr}\eqno(5.12)$$
where the dots stand for the other terms of higher order in the (anti)ghosts,
while the coefficients $a_1$, $a_2$, $a_4$ and $c$ are given by eqs.~(2.6)
and (2.7), respectively.

Eq.~(5.12) immediately yields the BRST charge nilpotency conditions:
$$\eqalign{
c_{\rm tot}~&\equiv c +c_{\rm gh}~=~\fracmm{k(N^2+6k-10)}{2(N+k-3)}-N^2+12N-26
=0~,\cr
s_{\rm tot}~&\equiv ~a_1 + (a_1)_{\rm gh}~=~\fracmm{k(N+2k-4)}{N+k-3}
 -\fracmm{k(N-1)(N-2)}{2(N+k-3)}\cr
{}~&~\qquad\qquad\qquad~~~~~ -\fracmm{4\h(N-1)}{N+k-3} +2 = 0~,\cr
k_{\rm tot}~&\equiv ~k + k_{\rm gh}~=~k+2N-6=0~,\cr
&~\qquad\fracmm{ \h(N+2k-4)}{N+k-3}-\fracmm{N-2}{2(N+k-3)}~ = 0~.\cr}
\eqno(5.13)$$

The first line of eq.~(5.13) just means the vanishing total central charge,
where
the value of $c_{\rm gh}$ is dictated by the standard formula of conformal
field theory \cite{book}
$$\eqalign{
c_{\rm gh}& = 2\sum_{\l} n_{\l}(-1)^{2\l+1}\left(6\l^2-6\l+1\right)\cr
& = 1\times (-26) + N\times (+11) + \ha N(N-1)\times (-2) = -26 +12N - N^2~,
\cr}\eqno(5.14)$$
 $\l$ is conformal dimension and $n_{\l}$ is a number of the conjugated
ghost pairs: $\l=2,3/2,1$ and $n_{\l}=1,N,\ha N(N-1)$, respectively. The zero
central charge condition alone has two solutions,
$$\eqalign{
k~=~&~6-2N~,\cr
6k~=~&~N^2-12N+26~,\cr}
\eqno(5.15)$$
but only the first of them is compatible with the third equation (5.13). The
fourth equation (5.13) just determines the renormalisation parameter $\h$.
Finally, the second equation (5.13) can be interpreted as the vanishing total
supersymmetric anomaly. Since the supersymmetry is non-linearly realised, this
anomaly does not have to vanish as a consequence of the other equations (5.13),
but restricts $N$ as the only remaining parameter. Substituting
$$k=6-2N~,\qquad {\rm and }\qquad \h=\fracmm{N-2}{2(8-3N)}~,\eqno(5.16)$$
into the second equation (5.13), we find
$$6(N-3)+\fracmm{(N-1)(N-2)(N-5)}{N-3}=0~,\eqno(5.17)$$
which has only one solution, $N=4$. Therefore, though the system of four
equations (5.13) for only three parameters $\h$, $k$ and $N$ is clearly
overdetermined (while $N$ is a positive integer!), there is still the only
solution, namely
$$N=4~,\qquad k=-2~,\qquad \h=-\fracmm{1}{4}~.\eqno(5.18)$$

The nilpotent quantum BRST charge for the Bershadsky-Knizhnik $SO(N)$-extended
 QSCAs exists, therefore, only when $N=4$. The corresponding BRST current reads
 $$\eqalign{
j_{\rm BRST}(z)=~&~ cT +\g^i G^i + \tilde{c}^a J^a +bc\pa c
-c\tilde{b}^a\pa\tilde{c}^a
 -\ha c\g^i\pa\b^i-\fracm{3}{2}c\b^i\pa\g^i \cr
& -b\g^i\g^i+\tilde{b}^a (t^a)^{ij}\left(\g^i\pa\g^j-\g^j\pa\g^i\right)
-\tilde{c}^a(t^a)^{ij}\b^i\g^j \cr
& -\ha f^{abc}\tilde{c}^a\tilde{c}^b\tilde{b}^c
 -\left[\left(t^at^b+t^bt^a\right)^{ij}+2\d^{ab}\d^{ij}\right]
J^a\tilde{b}^b\g^i\g^j \cr
& -\fracmm{1}{6}\left[\left(t^at^b+t^bt^a\right)^{ij}+2\d^{ab}\d^{ij}\right]
\left[\left(t^ct^d+t^dt^c\right)^{kl}+2\d^{cd}\d^{kl}\right]\times\cr
& \times f^{ace}
\tilde{b}^b\tilde{b}^d\tilde{b}^e\g^i\g^j\g^k\g^l~.\cr}\eqno(5.19)$$

We are now in a position to consider a construction of the quantum BRST charge
for the $\hat{D}(1,2;\a)$ QSCA generalising the $SO(4)$-based
Bershadsky-Knizhnik QSCA at $N=4$. The (anti)commutation relations of the
$\hat{D}(1,2;\a)$ QSCA were given in sect.~3. The ghost/antighost fields for
the internal $\Hat{su(2)}\oplus\Hat{su(2)}$ KM component of the
$\hat{D}(1,2;\a)$
QSCA are now denoted by $\tilde{c}^{aA}(z),~\tilde{b}^{aA}(z)$,
where $A=+,-\;$.

The natural ansatz for the quantum BRST charge of the $\hat{D}(1,2;\a)$ QSCA
is given by (see Appendix)
$$\eqalign{
Q_{\rm BRST}=~&~ c_{-n}L_n + \g^i_{-r}G^i_r + \tilde{c}^{aA}_{-n}J^{aA}_n
-\ha (m-n)c_{-m}c_{-n}b_{m+n}
+ nc_{-m}\tilde{c}^{aA}_{-n}\tilde{b}^{aA}_{m+n}\cr
& +\left(\fracm{m}{2}-r\right)c_{-m}\b^i_{m+r}\g^i_{-r}
-b_{r+s}\g^i_{-r}\g^i_{-s}
-\ha\tilde{c}^{aA}_{-m}(t^{aA})^{ij}\b^i_{m+r}\g^j_{-r}\cr
& + \h_2 b_2 (r-s) \tilde{b}^{a+}_{r+s}(t^{a+})^{ij}\g^i_{-r}\g^j_{-s}
  + \h_3 b_3 (r-s) \tilde{b}^{a-}_{r+s}(t^{a-})^{ij}\g^i_{-r}\g^j_{-s} \cr
& -\ha \ve^{abc}\tilde{c}^{a+}_{-m}\tilde{c}^{b+}_{-n}\tilde{b}^{c+}_{m+n}
 -\ha \ve^{abc}\tilde{c}^{a-}_{-m}\tilde{c}^{b-}_{-n}\tilde{b}^{c-}_{m+n}
 -\fracmm{1}{2}b_4\L^{ij}_{aAbB}J^{aA}_{r+s+m}\tilde{b}^{bB}_{-m}
\g^i_{-r}\g^j_{-s} \cr
&   -\fracmm{1}{24}b^2_4\L^{ij}_{aAbB}\L^{kl}_{cAdD}
 \ve^{ace}\d_{m+n+p,r+s+t+u}\tilde{b}^{bB}_m\tilde{b}^{dD}_n
(\tilde{b}^{e+}_p+\tilde{b}^{e-}_p)
\g^i_{-r}\g^j_{-s}\g^k_{-t}\g^l_{-u}~,\cr}\eqno(5.20a)$$
or, equivalently,
$$\eqalign{
j_{\rm BRST}(z)=~&~ cT + \g^iG^i + \tilde{c}^{aA}J^{aA} + bc\pa c
- c\tilde{b}^{aA}\pa\tilde{c}^{aA}
-\ha c\g^i\pa\b^i-\fracm{3}{2}c\b^i\pa\g^i  -b\g^i\g^i \cr
&~ -\ha\tilde{c}^{aA}(t^{aA})^{ij}\b^i\g^j
- \left[\h_2 b_2 \tilde{b}^{a+}(t^{a+})^{ij}
  + \h_3 b_3 \tilde{b}^{a-}(t^{a-})^{ij}\right](\g^i\pa\g^j-\g^j\pa\g^i) \cr
& -\ha \ve^{abc}\tilde{c}^{a+}\tilde{c}^{b+}\tilde{b}^{c+}
 -\ha \ve^{abc}\tilde{c}^{a-}\tilde{c}^{b-}\tilde{b}^{c-}
 -\fracmm{1}{2}b_4\L^{ij}_{aAbB}J^{aA}\tilde{b}^{bB}\g^i\g^j \cr
&   -\fracmm{1}{24}b^2_4\L^{ij}_{aAbB}\L^{kl}_{cAdD}
 \ve^{ace}\tilde{b}^{bB}\tilde{b}^{dD}(\tilde{b}^{e+}+\tilde{b}^{e-})
\g^i\g^j\g^k\g^l~,\cr}\eqno(5.20b)$$
where two quantum renormalisation parameters $\h_2$ and $\h_3$ have been
introduced, and $\L^{ij}_{aAbB}$ denote the $\hat{D}(1,2;\a)$ QSCA 4-point
 `structure constants', see eq.~(4.3).

Requiring the quantum BRST charge (5.20) to be nilpotent, yields the following
equations:
\begin{itemize}
\item from the terms $c(z)c(w)/(z-w)^4\,$:
$$c_{\rm tot}\equiv c+c_{\rm gh}=\left[\fracmm{6(k^++1)(k^-+1)}{k^++k^-+2}-3
\right]+6=0~,\eqno(5.21a)$$
where the central charge $c$ is now given by eq.~(3.9) and $c_{\rm gh}=+6$
according to eq.~(5.14);
\item from the terms $\g^i(z)\g^i(w)/(z-w)^3\,$:
$$s_{\rm tot}~\equiv ~b_1 + (b_1)_{\rm gh}~=~ b_1 +\fracmm{3}{2}b_4(k^++k^-)
-6(\h_2b_2+\h_3b_3) +2=0~,\eqno(5.21b)$$
where the parameters $b_1$, $b_2$ and $b_4$ are given by eq.~(3.8);
\item  from the terms $\tilde{c}^{a\pm}(z)\tilde{c}^{a\pm}(w)/(z-w)^2\,$:
$$k^{\pm}_{\rm tot}~\equiv ~k^{\pm} + 2 =0~,\eqno(5.21c)$$
\item from the terms $J^{a\pm}(t^{a\pm})^{ij}\g^i\pa\g^j/(z-w)\,$:
$$-2\h_2b_2-2b_4=-2\h_3b_3-2b_4=0~.\eqno(5.21d)$$
\end{itemize}

Note that eqs.~(5.13) and (5.21) are consistent with each other in the case of
$N=4$ and
$$k= k^+=k^-=-2~,\eqno(5.22)$$
as they should. It provides a good check of our calculations. Moreover,
eq.~(5.22) is, in fact, the {\it only} consistent solution to eq.~(5.21).
This means that the BRST quantisation of the non-linear  $\hat{D}(1,2;\a)$
QSCA can only be consistent if both its $\Hat{su(2)}$ KM components enter
symmetrically, i.e. when this quantum non-linear algebra is actually the
$SO(4)$-based Bershadsky-Knizhnik QSCA, with $k=-2$ and $c=3k=-6$. This is to
be compared with the known fact \cite{mm} that the quantum BRST charge for the
 `small' $N=4$ SCA, whose all central terms are related and proportional to
central charge, is only nilpotent when $c=-12$ (see also Table I).

A connection between the non-linear $SO(4)$-based Bershadsky-Knizhnik QSCA and
the `small' linear $SU(2)$-based SCA exists via the linearisation of the
former into the `large' linear $SU(2)\otimes SU(2)\otimes U(1)$-based SCA and
taking the limit either $k^+\to 0$ or $k^-\to 0$ (see sect.~3). Since (i)
there is no nilpotent QSCA BRST charge for the case of $k^+\neq k^-$, and (ii)
it does not make sense to gauge and BRST quantise {\it all} generators of
the `large' $N=4$ linear SCA, there seems to be no direct connection
between the nilpotent BRST operators for these two $N=4$ (Q)SCAs.
\vglue.2in

\section{Conclusion}

In our paper we constructed the quantum BRST charges for the {\it orthogonal}
(i.e. with the $\Hat{so(N)}$ KM component) series of Bershadsky-Knizhnik
non-linear QSCAs and for the particular quantum $\hat{D}(2,1;\a)$ QSCA
generalising them at $N=4$. We found only one nilpotent solution, namely, when
 $N=4$, $k=-2$ and $\a=1$. Eqs.~(5.18) and (5.22) constitute our main results.
They are apparently in line with the
analogous fact \cite{ssvn} that the BRST quantisation for all {\it unitary}
(i.e. with the $\Hat{u(N)}$ KM component) Bershadsky-Knizhnik QSCAs breaks
down for $N\geq 3$, since their
 BRST charge nilpotency conditions are always in conflict with the Jacobi
identities.~\footnote{The unitary series of Bershadsky-Knizhnik QSCAs do {\it
not} admit unitary representations for $N\geq 3$ also\cite{sch3}.}

The existence of the nilpotent quantum BRST operator for the non-linear
$SO(4)$-based Bershadsky-Knizhnik QSCA, and the existence of the corresponding
$W$-type $N=4$ conformal supergravity in two dimensions, imply the existence
of a new $N=4$ supersymmetric $W$-type string theory. The 2d supergravity
field equations of motion, which follow from the action (4.6) or (4.7), impose
the vanishing of the corresponding currents. At the quantum level, these
conditions can be interpreted, as in ordinary string theories, as operator
constraints on physical states. By interpreting zero modes of scalar fields in
matter QSCA realisations as spacetime coordinates, one arrives at a
first-quantised description of string oscillations. The structure of a matter
 part of the new string theory action will be discussed elsewhere \cite{gaket}.

Gauging the local symmetries of the $SO(4)$-based Bershadsky-Knizhnik QSCA
results in the positive total ghost central charge contribution,
  $c_{\rm gh}=6$. When adding the matter $(\j^i,\f)$ to linearise this algebra
(see sect.~3), one adds $+3$ to the total central charge. In addition, the
anomaly-free solution requires $k=-2<0$. Therefore, there is no way to build
an anomaly-free string theory by using only unitary representations.
For example, in the `basic' $c=3$ unitary non-linear representation (sect.~3),
the matter represented by $(\j^i,\f)$ is unified with the currents
$(T_{\rm tot},\,G^i_{\rm tot},\,J_{\rm tot}^{a\pm})$ into
one $N=4$ {\it linear} supermultiplet with $8_{\rm B}\oplus 8_{\rm F}$
components. As was suggested in ref.~\cite{sch1}, one may try to interpret the
only bosonic coordinate $\f$ there as a string world-sheet coordinate, with
the non-linearly realised world-sheet supersymmetry. However, it would then be
impossible to match the anomaly-free condition.

When choosing, instead, a non-unitary representation of the $SO(4)$-based QSCA
with  $k=-2$, one can get the desired anomaly-free matter contribution,
$c_{\rm m}=-6$, but then a space-time interpretation and a physical
significance of the construction, if any, become obscure. Despite of
all this, we  believe that it is worthy to know how many string models,
consistent from the mathematical point of view, can be constructed at all.
Relying on the argument based on the existence of a gauge-invariant action and
 a nilpotent BRST operator, our answer reads: at $N=4$ there are only two
different critical string theories within the requirements mentioned in the
Introduction.

Our results could still be generalised. In addition to the critical $N=2$
extended fermionic strings, based on {\it complex} numbers and the linear
$N=2$ SCA, and the critical $N=4$ extended fermionic strings, based on {\it
quaternionic} numbers and on either the linear `small' $N=4$ SCA or the
non-linear $SO(4)$-based Bershadsky-Knizhnik QSCA, new consistent string
theories, based on {\it octonionic} numbers and having $N=7$ or $N=8$
world-sheet supersymmetry, may exist. The exceptional $N=7$ and $N=8$ QSCAs
are known to arise via Drinfeld-Sokolov-type reduction from affine versions of
the exceptional Lie superalgebras $G(3)$ and $F(4)$ \cite{fl,bo}. Their
central charges are
$$c_7 = \fracmm{k(9k+31)}{2(k+3)}~,\qquad c_8 = \fracmm{2k(2k+11)}{k+4}~,$$
where $k$ is  an arbitrary `level' of the corresponding KM subalgebra. Those
exceptional QSCAs are indeed related with octonions \cite{fl,bo}, and may even
be identified with the octonionic generalisations of the $SO(3)$- and
$SO(4)$-based Bershadsky-Knizhnik QSCAs, respectively.

Like the $W$ algebras and unlike the linear SCAs, the non-linear QSCAs do not
yet have a natural geometrical interpretation, which is yet another
obstruction for their applications in physics.
\vglue.2in

\noindent{\Large\bf Acknowledgements}
\vglue.2in
\noindent
I acknowledge discussions with S.~J.~Gates Jr., J. W. van Holten
 and O.~Lechtenfeld.

\newpage

\noindent{\Large\bf Appendix: BRST Charge for Non-Linear Algebras}
\vglue.2in

In this Appendix we describe the procedure to obtain a nilpotent quantum
BRST operator $Q_{\rm BRST}$ directly from a quadratically non-linear operator
algebra. We first show how to construct a classical BRST charge $Q$, satisfying
the classical `master equation' $\{Q,Q\}_{\rm P.B.}=0$, for an arbitrary
classical (quadratically) non-linear algebra and, then, how to `renormalise'
the naively quantised operator $Q$ to a nilpotent quantum BRST charge
 $Q_{\rm BRST}$, satisfying the quantum `master eqation' $Q^2_{\rm BRST}=0$.

In classical theory, an algorithm for the construction of a BRST operator $Q$
is known due to Fradkin and Fradkina \cite{ff}. Consider a set of bosonic
generators $B_i$ and fermionic generators $F_{\a}$, which satisfy a graded
non-linear algebra of the form
$$\eqalign{
\{B_i,B_j\}_{\rm P.B.}=~&~f_{ij}{}^kB_k~,\cr
\{B_i,F_{\a}\}_{\rm P.B.}=~&~f_{i\a}{}^{\b}F_{\b}~,\cr
\{F_{\a},F_{\b}\}_{\rm P.B.}=~&~f_{\a\b}{}^{i}B_{i}+\L_{\a\b}{}^{ij}B_iB_j~,
\cr}\eqno(A.1)$$
in terms of the graded Poisson (or Dirac) brackets, with some 3-point and
4-point `structure constants', $f_{ij}{}^k$, $f_{i\a}{}^{\b}$, $f_{\a\b}{}^{i}$
 and $\L_{\a\b}{}^{ij}$, respectively, which have to be ordinary numbers. The
symmetry properties of these constants with respect to exchanging their indices
 obviously follow from their definition by eq.~(A.1), and they are going to be
 implicitly assumed below. When using the unified index notation,
$A\equiv(i,\a),\ldots~$, the Jacobi identities for the classical graded algebra
 of eq.~(A.1) take the form
$$\eqalign{
f_{[AB}{}^{D}f_{C\}D}{}^E~=~&~0~,\cr
\L_{[AB}{}^{DE}f_{C\}D}{}^F +\L_{[AB}{}^{DF}f_{C\}}{}^E
+f_{[AB}{}^D\L_{C\}D}{}^{EF}~=~&~0~.\cr}\eqno(A.2)$$
As is clear from eq.~(A.2), $f_{AB}{}^C$ are to be the structure constants of
a graded Lie algebra.~\footnote{We assume that all symmetry operations with
unified indices also have to be understood in the graded sense. In particular,
a graded `antisymmetrisation' of indices with unit weight (denoted by mixed
brackets $\[\;\}$ here) actually means the antisymmetrisation for
bosonic-bosonic or bosonic-fermionic index pairs, but the symmetrisation for
indices which are both fermionic.}

According to the classical BRST procedure,~\footnote{See, e.g., ref.~\cite{hen}
for a review.} one introduces an anticommuting ghost-antighost pair
$(c^m,b_m)$ for each of the bosonic generators $B_m$, and the commuting
ghost-antighost pair $(\g^{\a},\b_{\a})$ for each of the fermionic generators
$F_{\a}$. The ghosts satisfy (graded) bracket relations
$$\eqalign{
\{c^m,c^n\}_{\rm P.B.}=\{b_m,b_n\}_{\rm P.B.}=0~,~&~\{c^m,b_n\}_{\rm P.B.}=
\d\ud{m}{n}~,\cr
\{\g^{\a},\g^{\b}\}_{\rm P.B.}=\{\b_{\a},\b_{\b}\}_{\rm P.B.}=0~,~&~
\{\g^{\a},\b_{\b}\}_{\rm P.B.}=\d\ud{\a}{\b}~.\cr}\eqno(A.3)$$

Additional ghosts for the composite generators $B^iB^j$ are not needed since
invariance of the classical theory under $B^i$ already implies
invariance under $B^iB^j$ \cite{ff}.

The classial BRST charge $Q$ is  given by \cite{ff}
$$\eqalign{
Q = ~&~ c^nB_n + \g^{\a}F_{\a} +\ha f_{ij}{}^kb_kc^jc^i
+ f_{i\a}{}^{\b}\b_{\b}\g^{\a}c^i -\ha f_{\a\b}{}^nb_n\g^{\b}\g^{\a} \cr
&  -\ha \L_{\a\b}{}^{ij}B_ib_j\g^{a}\g^{\b}
-\fracm{1}{24}\L_{\a\b}{}^{ij}\L_{\g\d}{}^{kl}f_{ik}{}^{m}
b_jb_lb_m\g^{\a}\g^{\b}\g^{\g}\g^{\d}~.\cr}\eqno(A.4)$$
Compared to the standard expression for the linear algebras $(\L=0)$, the
BRST charge in eq.~(A.4) has the additional $3$-(anti)ghost terms, dependent on
the initial bosonic generators $B_i$, and the $7$-(anti)ghost terms as well.
It is easy to check that the classical `master equation'
$$\{Q,Q\}_{\rm P.B.}=0\eqno(A.5)$$
follows from eq.~(A.2) and the related identity
$$\L_{ \{\a\b}{}^{ij}\L_{\g\d\} }{}^{kl}f_{ik}{}^{m}=
\L_{ \{\a\b}{}^{i[j}\L_{\g\d\} }{}^{|k|l}f_{ik}{}^{m]}~.\eqno(A.6)$$

The graded classical algebra (A.1) can be extended to include (classical)
central extensions. However, the condition (A.5) forces them to vanish
\cite{ssvn}. This explains why we have not introduced central extensions for
the classical $\tilde{D}_4$ algebra in sect.~4.

The classical BRST charge (A.4) may serve as the starting point in a
construction of nilpotent quantum BRST charge $Q_{\rm BRST}$ associated with
the corresponding graded non-linear quantum algebra \cite{ssvn}. Since we are
actually interested in quantizing QSCAs, we can assume that all operators are
just currents, with a holomorphic dependence on $z$. In other words,  let
generators of the non-linear algebra under consideration carry an additional
(affine) index (see eq.~(2.1), for example). In particular, in eq.~(A.3) one
should simply replace the (graded) Poisson brackets by (anti)commutators,
$$\eqalign{
\[c^m,c^n\]=\[b_m,b_n\]=0~,~&~\[c^m,b_n\]= \d\ud{m}{n}~,\cr
\{\g^{\a},\g^{\b}\}=\{\b_{\a},\b_{\b}\}=0~,~&~
\{\g^{\a},\b_{\b}\}=\d\ud{\a}{\b}~,\cr}\eqno(A.7)$$
assuming that the indices $n$ and $\a$ are, in fact, multi-indices $nn'$  and
$\a r'$, where $n'$ and $r'$ are affine indices (of Fourier modes) while $n$
and $\a$ count different bosonic and fermionic currents, respectively.

In addition, in quantum theory, one must take into account central extensions
and the normal ordering needed for defining products of bosonic generators.
This results in the quantum (anti)commutation relations
$$\eqalign{
\[B_i,B_j\]=~&~f_{ij}{}^kB_k +h_{ij}Z~,\cr
\[B_i,F_{\a}\]=~&~f_{i\a}{}^{\b}F_{\b}~,\cr
\{F_{\a},F_{\b}\}=~&~h_{\a\b}Z+f_{\a\b}{}^{i}B_{i}+\L_{\a\b}{}^{ij}:B_iB_j:~,
\cr}\eqno(A.8)$$
where the central charge generator $Z$ commutes with all the other generators,
and the new constants $h_{ij}$ and $h_{\a\b}$ are supposed to be restricted
by the Jacobi identities (see sects.~2 and 3). The ghost/anti-ghosts become
operators, and they also require normal ordering for their products. Although
no general procedure seems to exist, which would explain how to fully
 `renormalise' the naively quantised (i.e. only normally-ordered) charge $Q$
to a nilpotent quantum-mechanical
 operator $Q_{\rm BRST}$, the answer is known for a particular class of
quantum algebras of the $W$-type due to refs.~\cite{tm,ssvn}. The quantum
algebra $\hat{D}(1,2;\a)$ falls into this class, so that we may expect that,
similarly to the quantum $W$ algebras considered in refs.~\cite{tm,ssvn}, the
only non-trivial modification of eq.~(A.4) in quantum theory amounts to a
{\it multiplicative} renormalisation of the structure constants $f_{\a\b}{}^i$,
$$f_{\a\b}{}^i\to (f_{\rm ren})_{\a\b}{}^i\equiv \h f_{\a\b}{}^i~,\eqno(A.9)$$
after the formal replacement of the graded Poisson brackets by
(anti)commutators and the normal ordering, namely
$$\eqalign{
Q_{\rm BRST} = ~&~ c^nB_n + \g^{\a}F_{\a} +\ha f_{ij}{}^k:b_kc^jc^i:
+ f_{i\a}{}^{\b}:\b_{\b}\g^{\a}:c^i
-\ha\h f_{\a\b}{}^nb_n\g^{\b}\g^{\a} \cr
&  -\ha \L_{\a\b}{}^{ij}B_ib_j\g^{a}\g^{\b}
-\fracm{1}{24}\L_{\a\b}{}^{ij}\L_{\g\d}{}^{kl}f_{ik}{}^{m}
b_jb_lb_m\g^{\a}\g^{\b}\g^{\g}\g^{\d}~.\cr}\eqno(A.10)$$
This {\it ansatz} for the quantum BRST operator introduces only a few
(normally, only one) additional renormalization parameters $\h$ to be
determined from the nilpotency condition.
Since the central extension parameters of the quantum non-linear algebra are
severely restricted by the Jacobi identities, whereas the quantum BRST
charge nilpotency conditions normally lead to some more restrictions on their
values, the procedure could make the parameters to be overdetermined, in
general. Therefore, an existence of the quantum BRST charge is not guaranteed,
 and it is important to check consistency in each particular case.

Eq.~(A.10) was used in the text (sect.~5), in order to write down the
{\it ans\"atze} (5.8) and (5.20)
 for a quantum BRST operator $Q_{\rm BRST}$ in the cases of Bershadsky-Knizhnik
orthogonal series of non-linear algebras and $\hat{D}(2,1;\a)$ QSCA,
respectively. Since there was {\it \'a priori} no guarantee that the
quantum BRST charges constructed this way are going to be nilpotent, it is not
very surprising that the whole construction turns out to be consistent only
when $N=4$ and $\a=1$, i.e. for the orthogonal $SO(4)$-based
Bershadsky-Knizhnik quantum  quasi-superconformal algebra alone (sect.~5).
 \vglue.2in

\end{document}